\documentclass[twocolumn]{apex}
\usepackage{graphicx}

\title{Epitaxial Growth of Superconducting Ba(Fe$_{1-x}$Co$_x$)$_2$As$_2$ Thin Films on Technical Ion Beam Assisted Deposition MgO Substrates}

\author{Kazumasa\,Iida, Jens\,H\"{a}nisch, Sascha\,Trommler, Vladimir\,Matias$^1$, Silvia\,Haindl, Fritz\,Kurth, Irene\,Lucas del Pozo, Ruben\,H\"{u}hne, Martin\,Kidszun, Jan\,Engelmann, Ludwig\,Schultz, and Bernhard\,Holzapfel}
\inst{Leibniz-Institut f\"{u}r Festk\"{o}rper-und Werkstoffforschung (IFW) Dresden, P.\,O.\,Box 270116, 01171 Dresden, Germany\\
$^1$Superconductivity Technology Center, Los Alamos National Laboratory, Los Alamos, NM 87545, U.S.A. }

\abst{The biaxially textured growth of superconducting Co-doped BaFe$_2$As$_2$ (Ba-122) thin films has been realized on ion beam assisted deposition (IBAD) MgO coated conductor templates by employing an iron buffer architecture. The iron pnictide coated conductor showed a superconducting transition temperature of 21.5\,K, which is slightly lower than that of Co-doped Ba-122 films on single crystalline MgO substrates. A self-field critical current density of over 10$^{5}$\,A$\cdot$cm$^{-2}$ has already been achieved even at 8\,K. The current experiment highlights the potential of possible coated conductor applications of the iron-based superconductors.}

\begin{document}
\maketitle
As recently demonstrated, the implementation of an Fe buffer layer is beneficial for epitaxial growth of Co-doped BaFe$_2$As$_2$ (Ba-122) thin films on MgO single crystalline substrates since the metallic bond between the Fe layer and the Co-doped Ba-122 takes place on the Fe sublayer within the FeAs tetrahedron.\cite{Tom}
Furthermore, a small lattice mismatch of 2.3\% is realized between the (001) surface plane of Fe, which is rotated 45$^\circ$ in-plane, and the Fe sublayer in the Co-doped Ba-122 unit cell. As a result, excellent superconducting properties have been realized in Fe/Ba-122 bilayer system.\cite{Kazu-1} This opens the opportunity to grow Co-doped Ba-122 films on technical substrates (e.g., Hastelloy), on which textured MgO templates are prepared by ion beam assisted deposition (IBAD).
 
IBAD offers biaxially textured buffer layers for epitaxial growth of functional materials, such as high-$T_{\rm c}$ superconducting YBa$_2$Cu$_3$O$_7$ (YBCO).\cite{Iijima} In particular, the IBAD-MgO templates on Hastelloy are used for the 2$^{nd}$ generation YBCO coated conductor as the texture nucleation offers significantly faster processing than that in the case of using other materials.\cite{IBAD-MgO} At low temperatures, Co-doped Ba-122 offered the opportunity of very high upper critical field in combination with a much lower anisotropy compared to YBCO, which is one of the advantages for coated conductor applications. However, the critical current density ($J_{\rm c}$) values of Fe-based superconducting wires or tapes published to date are not practical level, showing typically less than 10$^{3}$\,A$\cdot$cm$^{-2}$ even at low temperature.\cite{wire-1, wire-2} 

Similarly to YBCO,  grain boundaries (GBs) with misorientation angles above $6^{\circ}$ seriously reduce the critical current for Co-doped Ba-122.\cite{Lee} The same strong weak-link behavior due to GBs has been reported in F-doped LaFeAsO polycrystalline thin films.\cite{Silvie} Hence, the current limiting effects across GBs in Fe-based superconductors have found a general consensus, necessitating biaxially textured growth to achieve high critical currents.

Here, we report on the crystalline quality and superconducting properties of Co-doped Ba-122 grown on IBAD-MgO Hastelloy, offering the possibility for coated conductor processing of the new Fe-based superconductors.

\begin{figure}
\begin{center}
\includegraphics[width=\columnwidth]{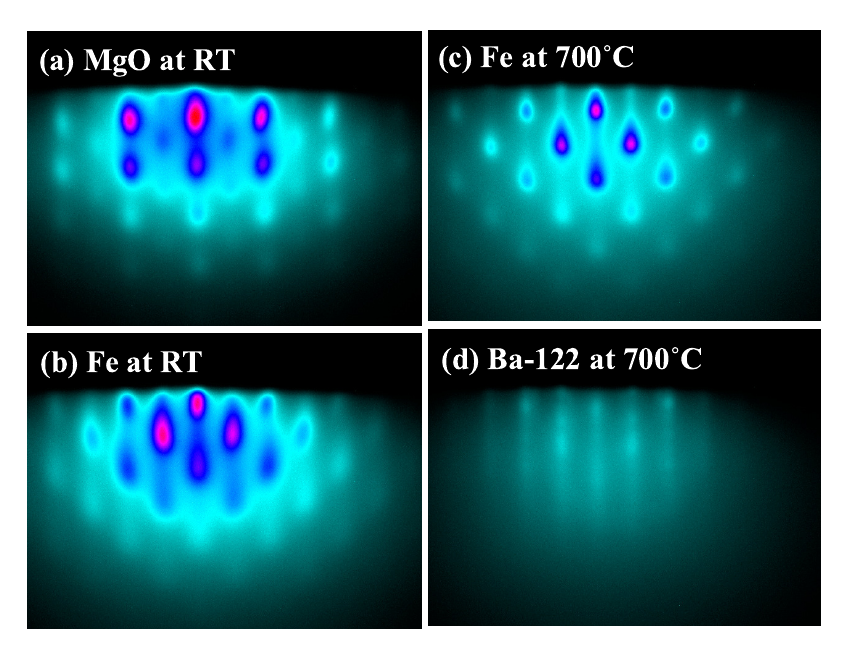}
\end{center}
\caption{(a) The RHEED image of an IBAD-MgO template at room temperature (RT) shows a good texture. The incident electron beam is parallel to [110] MgO. (b) Fe grows textured even at RT. (c) The crystalline quality improves at elevated temperatures. (d) Epitaxial growth of Co-doped Ba-122 on an Fe buffer layer.}
\label{fig:figure1}
\end{figure}

10\,mm wide commercial Hastelloy C-276 tapes were planarized by solution planarization deposition (SDP) with 15 layers of Y$_2$O$_3$ with a total thickness of 1\,$\mu$m.\cite{Matias-1} A roughness of around 0.6-0.8\,nm on $5\times5\,\mu$m$^2$ was achieved. The Y$_2$O$_3$ layer does not only provide a smooth and amorphous surface needed for IBAD but also protects the Ba-122 phase against detrimental diffusion of Ni and Cr from the Hastelloy tape. The IBAD-MgO layer, deposited at room temperature (RT), has a thickness of 5\,nm and is covered by a homoepitaxial MgO layer of around 40\,nm, which was deposited at $600^{\circ}$C [Fig.\,\ref{fig:figure1}(a)]. More details of the template preparation can be found in ref.\,\citeform{10}. For use in the pulsed laser deposition (PLD) system, the tape was mechanically cut into $10\times$10\,mm$^2$ pieces.

Fe layers of around 10\,nm were deposited at RT on IBAD-MgO Hastelloy substrates by PLD, using a KrF excimer laser (248\,nm) at a repetition rate of 8\,Hz in an ultra high vacuum chamber (base pressure of 10$^{-8}$\,mbar). Reflection high-energy electron diffraction (RHEED) confirmed the epitaxial growth of Fe buffer layers even at RT [Fig.\,\ref{fig:figure1}(b)], and their diffraction spots became sharper with increasing temperature as shown in Fig.\,\ref{fig:figure1}(c), indicating that the crystalline quality improves. The Fe-covered IBAD-MgO Hastelloy substrate was then heated to $700^{\circ}$C for the deposition of about 50\,nm thick Co-doped Ba-122 [Fig.\,\ref{fig:figure1}(d)]. The detailed PLD target preparation and deposition conditions can be found in ref.\,\citeform{11}.

\begin{figure}
\begin{center}
\includegraphics[width=\columnwidth]{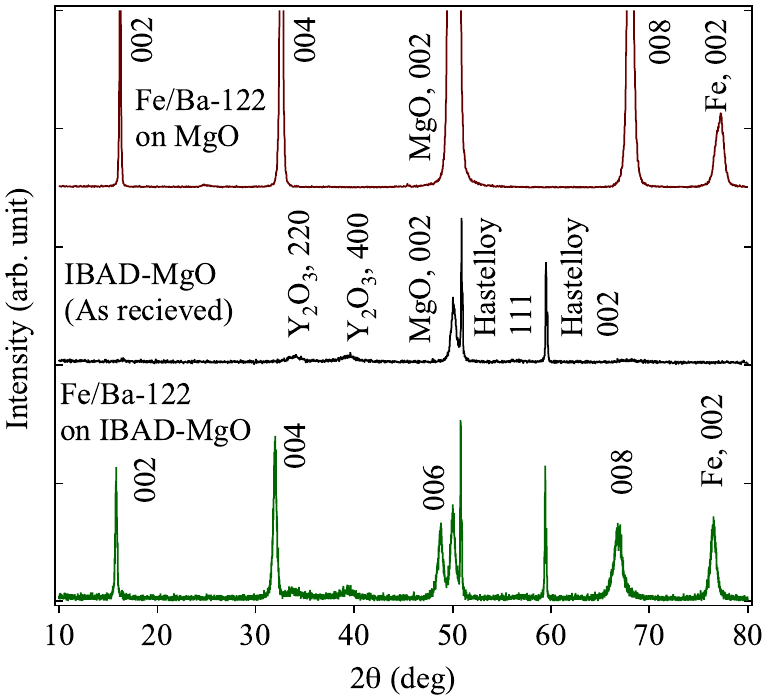}
\end{center}
\caption{The $\theta\rm/2\theta$-\,scans (Co-$K_\alpha$ radiation) of the Co-doped Ba-122 film on IBAD-MgO Hastelloy show a high phase purity and a $c$-axis texture (bottom traces). The data for Co-doped Ba-122 on a single crystalline MgO substrate and for bare IBAD-MgO Hastelloy are also plotted for clarity.}
\label{fig:figure2}
\end{figure}

Figure\,\ref{fig:figure2} exhibits the $\theta\rm/2\theta$-\,scans of the film on IBAD-MgO Hastelloy using Co-$K_\alpha$ radiation. All peaks were indexed with Co-doped Ba-122, Fe, Y$_2$O$_3$, MgO and Hastelloy, indicating a high phase purity. Major peaks observed in Fig.\,\ref{fig:figure2} were the 00$l$ reflection of the Co-doped Ba-122 and the 002 reflection of MgO together with Fe, implying that all deposited layers except Y$_2$O$_3$ were $c$-axis-oriented normal to the Hastelloy substrate plane. The Y$_2$O$_3$ peaks are due to a partial crystallization of the SDP Y$_2$O$_3$ layer during the deposition of the homoepitaxial MgO layer.

\begin{figure}
\begin{center}
\includegraphics[width=8cm]{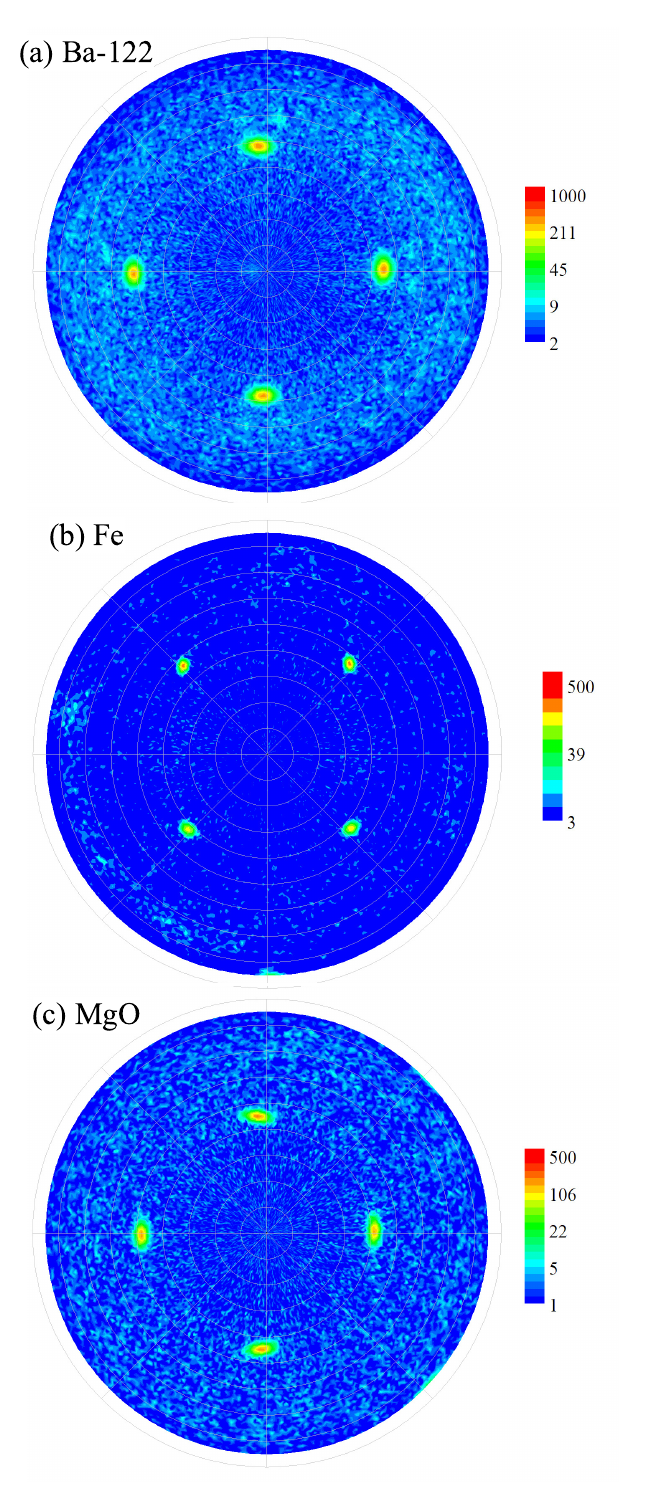}
\end{center}
\caption{(a) The 103 pole figure of Co-doped Ba-122 proves the in-plane texture. (b) Fe grows epitaxially on MgO with a $45^\circ$ rotation. (c) MgO is biaxially textured on Y$_2$O$_3$ bed layered Hastelloy. The epitaxial relation is (001)[100]Ba-122$\|$(001)[110]Fe$\|$(001)[100]MgO.}
\label{fig:figure3}
\end{figure}

In order to confirm the epitaxial relation of the stacking layers for Co-doped Ba-122, Fe and MgO, pole figure measurements were conducted using Cu-$K_\alpha$ radiation. The respective contour plots in the logarithmic scale for the 103 reflection of Co-doped Ba-122, the 110 reflection of Fe, and the 220 reflection of MgO show a clear four-fold symmetry, indicating that all the layers grow epitaxially with the relation (001)[100]Ba-122$\|$(001)[110]Fe$\|$(001)[100]MgO [Fig.\,\ref{fig:figure3}]. Table~\ref{tab:table1} shows a comparison of the crystalline quality between a film on IBAD-MgO and a Fe/Ba-122 bilayer on single crystalline MgO. The out-of-plane full width at half-maximum (FWHM) $\Delta\omega$, as well as the FWHM of the pole figures, $\Delta\phi$, of the Ba-122 film are similar to those of the underlaying Fe layer. Therefore, in both cases, the texture is transferred to the superconducting Ba-122 layer. In contrast to Fe/Ba-122 bilayers on single crystalline MgO substrates,\cite{Tom} the corresponding values are large, which is presumably due to the thin homoepitaxial layer of the IBAD-MgO templates. For the IBAD-MgO templates, the out-of-plane FWHM values of MgO reduce with increasing homo-epitaxial layer thickness and saturate of around 200 nm.\cite{10} The in-plane texture shows the same tendency, but without saturation up to micrometer thicknesses. Hence, it might be possible to further improve the crystalline quality of the Fe buffer layer by employing thicker IBAD-MgO templates, which leads to a higher degree of texture development in Co-doped Ba-122.

\begin{table}
\caption{Average FWHM values of the $\phi$\,-\,scans for MgO, Fe and Co-doped Ba-122 layers on IBAD-MgO Hastelloy. For comparison, the corresponding values on MgO single crystalline substrates are also shown. In both cases, the in-plane orientation of Co-doped Ba-122, $\Delta\phi_{\rm Ba-122}$, is almost the same as that of Fe, $\Delta\phi_{\rm Fe}$.}
\label{tab:table1}
\begin{tabular}{lccr}
\hline
Substrates&$\Delta\phi_{\rm MgO}$&$\Delta\phi_{\rm Fe}$&$\Delta\phi_{\rm Ba-122}$ \\
\hline
IBAD-MgO Hastelloy& $5.98^\circ$ & $4.71^\circ$ &  $5.13^\circ$ \\
MgO single crystal& $0.58^\circ$ & $1.05^\circ$ & $0.95^\circ$ \\
\hline
\end{tabular}
\end{table}

\begin{figure}
\begin{center}
\includegraphics[width=\columnwidth]{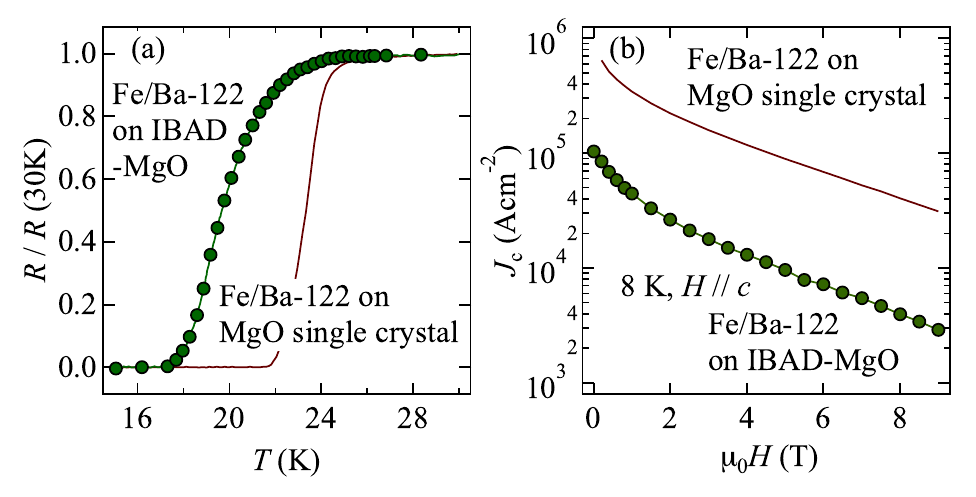}
\end{center}
\caption{(a) Normalized resistance of the Co-doped Ba-122 thin film on IBAD-MgO. The data were normalized to the value of 30\,K. For comparison, the film on a MgO single crystalline substrate is also plotted. (b) $J_{\rm c}-H$ characteristics for Co-doped Ba-122 thin film on IBAD-MgO at 8\,K. A self-field $J_{\rm c}$ of over 10$^{5}$\,A$\cdot$cm$^{-2}$ has been recorded.}
\label{fig:figure4}
\end{figure}

The normalized resistive traces of the Co-doped Ba-122 on IBAD-MgO Hastelloy are displayed in Fig.\,\ref{fig:figure4}(a). The measurement was conducted using a physical property measurement system (PPMS; Quantum Design) with a standard four-probe method. The onset superconducting transition temperature $T_{\rm c}$ of the film on IBAD-MgO Hastelloy was around 21\,K, which is slightly lower than that of the film on a MgO single crystalline substrate. As stated earlier, increased in-plane FWHM values indicate that GBs develop in the Co-doped Ba-122 layer, possibly leading to oxidation along the GBs, which would reduce $T_{\rm c}$. In addition, the broad transition width of around 3\,K may also be caused by the weak-link behavior. In YBCO, GBs reduce $T_{\rm c}$ even if they are not yet weak links (i.e., low-angle GBs) due to strain around the dislocation cores and the band bending effect. The same mechanisms might have a similar effect in the Co-doped Ba-122 film.

Albeit a relatively thick Y$_2$O$_3$ layer of $1\,\mu$m, an inter-diffusion of Ni and Cr from Hastelloy to the superconducting layer cannot be ruled out completely due to the porous structures of these Y$_2$O$_3$ layers. Since both transition elements can enter the Fe site in the Ba-122 lattice, this diffusion would result in either electron or hole doping. The former induces superconductivity with a maximum $T_{\rm c}$ of 19\,K,\cite{Ni} and the latter does not show any signs of superconductivity at all.\cite{Cr-doped} Nevertheless, the $T_{\rm c}$ of the resultant film on IBAD-MgO Hastelloy decreased only by a small amount compared with that of fully optimized films on MgO single crystalline substrates.

\begin{figure}
\begin{center}
\includegraphics[width=\columnwidth]{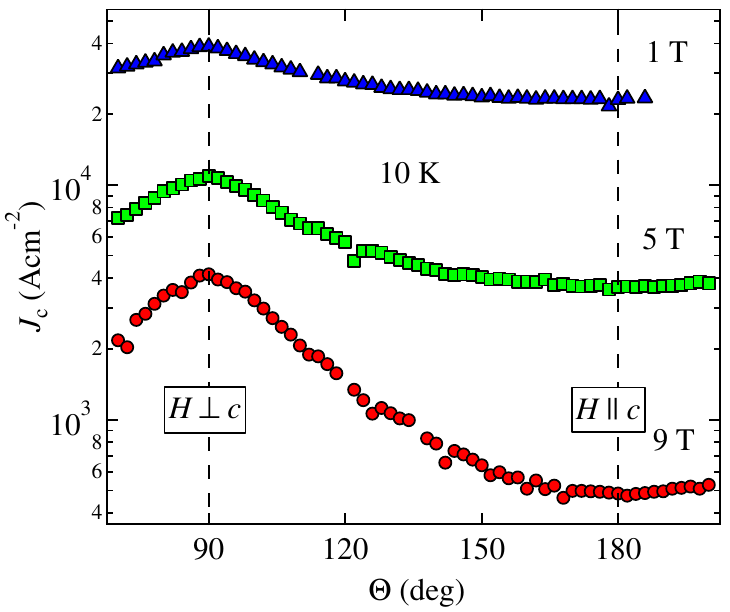}
\end{center}
\caption{Angular dependence of $J_{\rm c}$ for the Co-doped Ba-122 on IBAD-MgO substrates at 10\,K under several magnetic fields. The magnetic field $H$ was applied in the maximum Lorentz force configuration ($H$ perpendicular to $J$) at an angle of $\Theta$ measured from the $c$-axis.}
\label{fig:figure5}
\end{figure}

Shown in Fig.\,\ref{fig:figure4}(b) are $J_{\rm c}-H$ characteristics for a Co-doped Ba-122 thin film on IBAD-MgO at 8\,K. For these measurements, the films were cut into slabs measuring 1\,mm in width and 8\,mm in length with a wire saw. A criterion of 1\,$\rm\mu$V$\cdot$cm$^{-1}$ for evaluating $J_{\rm c}$ was employed. A self-field $J_{\rm c}$ of over 10$^{5}$\,A$\cdot$cm$^{-2}$ has already been achieved even at 8\,K. Further improvement in $J_{\rm c}-H$ performance is possible since the fully optimized Fe/Ba-122 bilayers on MgO single crystalline substrates show one order of magnitude higher $J_{\rm c}$ values than that of the film of this study. The angular-dependent critical current density $J_{\rm c}(\Theta)$ shown in Fig.\,\ref{fig:figure5} has a broad maximum at  $\Theta=90^{\circ}$, which arises from intrinsic pinning.\cite{Kazu-3}

In summary, we have demonstrated the biaxially textured growth of superconducting Co-doped Ba-122  thin films on IBAD-MgO Hastelloy by employing an iron buffer layer architecture. The film on IBAD-MgO Hastelloy showed a superconducting transition temperature of 21\,K with a large transition width of 3\,K. A self-field $J_{\rm c}$ of over 10$^{5}$\,A$\cdot$cm$^{-2}$ has already been achieved even at 8\,K. These results open the avenue to the possibility of coated conductor growth of the new iron pnictide superconductors.  

\acknowledgement
The authors thank M.\,K\"{u}hnel and U.\,Besold at Leibniz-Institut f\"{u}r Festk\"{o}rper-und Werkstoffforschung Dresden for their technical support. We also thank C.\,Sheehan at Los Alamos National Laboratory for supplying solution deposited substrates. This work was partially supported by the German Research Foundation.


\begin{thebibliography}{99}

\bibitem{Tom}
T. Thersleff, K. Iida, S. Haindl, M. Kidszun, D. Pohl, A. Hartmann, F. Kurth, J. H{\"a}nisch, R. H{\"u}hne, B. Rellinghaus, L. Schultz, and B. Holzapfel: Appl. Phys. Lett. {\bf97} (2010) 022506.
\bibitem{Kazu-1}
K. Iida, S. Haindl, T. Thersleff, J. H{\"a}nisch, F. Kurth, M. Kidszun, R. H{\"u}hne, I. M{\"o}nch, L. Schultz, B. Holzapfel, and R. Heller: Appl. Phys. Lett. {\bf97} (2010) 172507.
\bibitem{Iijima}
Y. Iijima, N. Tanabe, O. Kohno, and Y. Ikeno: Appl. Phys. Lett. {\bf60} (1992) 769.
\bibitem{IBAD-MgO}
C. P. Wang, K. B. Do, M. R. Beasley, T. H. Geballe, and R. H. Hammond: Appl. Phys. Lett. {\bf71} (1997) 2955.
\bibitem{wire-1}
Z. Gao, L. Wang, Y. Qi, D. Wang, X. Zhang, and Y. Ma: Supercond. Sci. Technol. {\bf21} (2008) 105024.
\bibitem{wire-2}
Y. Mizuguchi, K. Deguchi, S. Tsuda, T. Yamaguchi, H. Takeya, H. Kumakura, and Y. Takano: Appl. Phys. Express {\bf2} (2009) 083004.
\bibitem{Lee}
S. Lee, J. Jiang, J. D. Weiss, C. M. Folkman, C. W. Bark, C. Tarantini, A. Xu, D. Abraimov, A. Polyanskii, C. T. Nelson, Y. Zhang, S. H. Baek, H. W. Jang, A. Yamamoto, F. Kametani, X. Q. Pan, E. E. Hellstrom, A. Gurevich, C. B. Eom, and D. C. Larbalestier: Appl. Phys. Lett. {\bf95} (2009) 212505.
\bibitem{Silvie}
S. Haindl, M. Kidszun, A. Kauffmann, K. Nenkov, N. Kozlova, J. Freudenberger, T. Thersleff, J. H{\"a}nisch, J. Werner, E. Reich, L. Schultz, and B. Holzapfel: Phys. Rev. Lett. {\bf104} (2010) 077001.
\bibitem{Matias-1}
C. Sheehan, Y. Jung, T. Holesinger, D. M. Feldmann, V. Matias, C. Edney, J. F. Ihlefeld, and P. G. Clem: to be published in Appl. Phys. Lett.
\bibitem{10}
V. Matias, J. H{\"a}nisch, E. J. Rowley, and K. G{\"u}th: J. Mater. Res. {\bf24} (2009) 125.
\bibitem{11}
K. Iida, J. H{\"a}nisch, R. H{\"u}hne, F. Kurth, M. Kidszun, S. Haindl, J. Werner, L. Schultz, and B. Holzapfel:  Appl. Phys. Lett. {\bf95} (2009) 192501.
\bibitem{Ni}
N. Ni, A. Thaler, J. Q. Yan, A. Kracher, E. Colombier, S. L. Bud$'$ko, and P. C. Canfield: Phys. Rev. B {\bf82} (2010) 024519.
\bibitem{Cr-doped}
A. S. Sefat, D. J. Singh, L. H. VanBebber, Y. Mozharivskyj, M. A. McGuire, R. Jin, B. C. Sales, V. Keppens, and D. Mandrus: Phys. Rev. B {\bf79} (2009) 224524.
\bibitem{Kazu-3}
K. Iida, J. H{\"a}nisch, T. Thersleff, F. Kurth, M. Kidszun, S. Haindl, R. H{\"u}hne, L. Schultz, and B. Holzapfel:  Phys. Rev. B {\bf81} (2010) 100507(R).

\end{thebibliography}
\end{document}